\documentstyle[11pt,aaspp4]{article}
\newcommand{\asca}{{\sl ASCA }}

\newcommand{\sax}{{\sl SAX }}

\newcommand{\whipple}{{\sl Whipple }}

\newcommand{\rxte}{{\sl RXTE }}
\newcommand{\euve}{{\sl EUVE }}

\slugcomment{}

\lefthead{Takahashi et al.}

\righthead{Complex Spectral Variability of Mrk~421}

\begin{document}
\title{Complex Spectral Variability from Intensive Multi-wavelength
Monitoring of Mrk~421 in 1998}

\author{
T.~Takahashi\altaffilmark{1,2}, J.~Kataoka\altaffilmark{1,2}, 
G.~Madejski\altaffilmark{3,4},
J.~Mattox\altaffilmark{5}, 
C.M.~Urry\altaffilmark{6}, 
S.~Wagner\altaffilmark{7}, 
F.~Aharonian\altaffilmark{8},
M.~Catanese\altaffilmark{9},
L.~Chiappetti\altaffilmark{10},
P.~Coppi\altaffilmark{11},
B.~Degrange\altaffilmark{12}, 
G.~Fossati\altaffilmark{13},
H.~Kubo\altaffilmark{14},
H.~Krawczynski\altaffilmark{8},
F.~Makino\altaffilmark{1},
H.~Marshall\altaffilmark{15}, 
L.~Maraschi\altaffilmark{16}, 
F.~Piron\altaffilmark{12}, 
R.~Remillard\altaffilmark{15},
F.~Takahara\altaffilmark{17},
M.~Tashiro\altaffilmark{2}, 
H.~Terasranta\altaffilmark{18}, 
and T.~Weekes\altaffilmark{9}
}

\altaffiltext{1}{Institute of Space and Astronautical Science,
 Kanagawa 229-8510, Japan}
\altaffiltext{2}{Department of Physics, University of Tokyo,  
 Tokyo
 113-0033, Japan}
\altaffiltext{3} {Laboratory for High Energy Astrophysics, NASA/GSFC, 
	Greenbelt, MD 20771}
\altaffiltext{4} {Department of Astronomy, University of Maryland, Collge 
	Park, MD 20742}
\altaffiltext{5} {Astronomy Department, Boston University,  Boston, MA, 02215}
\altaffiltext{6} {Space Telescope Science Institute, 
	Baltimore, MD 21218}

\altaffiltext{7} {Landessternwarte Heidelberg, K\"{o}nigstuhl, D-69117 
	Heidelberg, Germany}
\altaffiltext{8} {Max-Planck-Institut F\"{u}r Kernphysic, 
	D-69029 Heidelberg, Germany}
\altaffiltext{9} {Fred Lawrence Whipple Observatory, Harvard-Smithsonian 
	CfA, Amado, AZ 85645}
\altaffiltext{10} {Instituto di Fisica Cosmica CNR, Milano, Italy}
\altaffiltext{11} {Department of Astronomy, Yale University, New Haven, CT 06520-8101}
\altaffiltext{12} {LPNHE Ecole Polytechnique Palaiseau, France},
\altaffiltext{13} {University of California at San Diego, 
 La Jolla, CA 92093-0424}
\altaffiltext{14} {Department of Physics, Tokyo Institute of Technology, 
	Tokyo, 152-8551,Japan}
\altaffiltext{15} {Center for Space Research, Cambridge, MA 02139}
\altaffiltext{16} {Osservatorio Astronomico di Brera, 20121 
	Milano, Italy.}
\altaffiltext{17} {Department of Earth and Space Science, Osaka University, Osaka, Japan}
\altaffiltext{18} {Mets\"ahove Radio Observatory, Kylmala, Finland}

\begin{abstract}

We conducted a multi-frequency campaign for the TeV blazar Mrk~421 in
1998 April.  The campaign started from a pronounced high amplitude 
flare recorded by \sax and $Whipple$;  \asca observation started three 
days later,
In the X--ray data, we detected multiple flares, occuring on 
time scales of about one day.  \asca data clearly reveal spectral 
variability.  The comparison of the data from \asca, \euve and \rxte indicates
that the  variability amplitudes in the low energy synchrotron component are
larger at  higher photon energies. In TeV $\gamma$--rays, large intra-day
variations  -- which were correlated with the X--ray flux -- were
observed when results from three Cherenkov telescopes are combined. The
RMS variability of TeV $\gamma$--rays was similar to that observed in hard
X--rays, above 10~keV. The X--ray light curve reveals flares which are 
almost symmetric for most of cases, implying the dominant time scale
is the light crossing time through the emitting region.  The structure
function analysis based on the continuous X--ray light curve of seven
days indicates that the characteristic time scale is $\sim$~0.5 day.  
The analysis of ASCA light curves in various energy bands appears to show
both soft (positive) and hard (negative) lags. These may not be real, as
systematic effects could also produce these lags, which are all much
smaller than an orbit.  If the lags of both signs are real, these imply
that the particle acceleration and X-ray cooling time scales are similar.

\end{abstract}


\section{Introduction}

According to well-established unified schemes for radio-loud AGN (Urry
\& Padovani 1995), blazars are distinguished by the fortuitous
alignment of their jets along the line of sight. This greatly enhances
the jet radiation, making them ideal sources for understanding jet
physics in all radio-loud AGN.
The 
required high energies of the radiating electrons indicate very efficient 
particle acceleration in the relativistic jets of these sources 
(Kubo et al. 1998; Ghisellini et al. 1998).

The BL Lac object Mrk~421 is unique as the first -- and so far, the
brightest -- High Frequency Peaked BL Lac objects (HBLs) with
$\gamma$--ray emission extending up to TeV energies, at a level
allowing detailed spectral and variability studies in the broadest
spectral range. The simple continuum spectra of Mrk~421 from radio to
UV and X--ray bands obtained previously imply that the emission in
these energy bands is due to charged particles radiating via the
synchrotron process (e.g., George, Warwick, \& Bromage 1988). The hard
X--rays and $\gamma$--rays from this object are likely to be produced
by inverse Compton scattering of synchrotron photons by the same
electrons (e.g., Ulrich, Maraschi, \& Urry 1997 and references
therein).  Judging from the results of previous TeV and
multi-wavelength monitoring campaigns of Mrk~421, correlations among
light curves at different energies (and, in particular, the pronounced
spectral evolution during flares seen in X--rays) have provided our
best opportunity to understand the high energy emission from blazar
jets (Macomb et al. 1995; Buckley et al. 1996; Takahashi et al. 1996;
Wagner 1996; Takahashi et al. 1999).  Specifically, the previous
(1994) campaigns found a soft lag of about 4000~s for soft X--ray (0.5
-- 1.0~keV) photons with respect to the hard X--ray (2 -- 7.5~keV)
band, which was interpreted as an effect of radiative cooling
(Takahashi et al. 1996).  However, the sparse sampling of previous
campaigns has prevented us from obtaining definitive conclusions.
Uninterrupted monitoring, lasting several times the characteristic
time scale, is the only way to investigate quantitatively the extreme
physical conditions at the sites of high-energy radiation.

Accordingly, we carried out a one-week continuous observation with
\asca, coordinated with \euve, and \sax. At the same time, TeV
detectors (CAT, HEGRA, and Whipple), optical telescopes, radio
antennae at 22~GHz and 37~GHz attempted to monitor the source every
night.  In this paper, we only present the most important result of
the multi-wavelength campaign.

\section{Observations and Results}

\subsection{Observations}

The essential information on the observations is summarized in Fig.~1,
which shows the normalized multi-wavelength (EUV, X--ray and TeV) 
light curves.
The logarithmic flux scale allows flares with equal amplitudes
to be represented by equal vertical excursions.  Complete details 
about the data analyses will be given in separate 
wavelength-based papers (Aharonian et al. 1999;   
Fossati et al. 2000a,b; Kataoka et al. 2000a); here we summarize the key
points.
 
Continuous coverage in X--rays with \asca lasted from 1998 April 23.97 to
April 30.8 UT, yielding a net exposure of $\sim$ 280~ks. 
 The results presented here are from the SIS detectors operated in
1-CCD mode because the count rate of the GIS detectors exceeded its
telemetry limit for 20 -- 30 \% of the observation time. The
background count rate and its fluctuations are negligible,
so we did not perform any background subtraction, so as to
avoid any artificial effects.  

\sax observations of Mrk~421 preceded the \asca observation by about three 
days (Maraschi et al.  1999). \rxte observed the source 
primarily only in low background orbits in order to minimize uncertainties 
in the background subtraction. 
EUVE observations covered the whole campaign, lasting
from April 19.4 to May 1.2 
The $\gamma$--ray observations at TeV energies were made with the CAT, HEGRA
and Whipple atmospheric Cherenkov telescopes. 
Optical observations with BVRI  filters were performed by the WEBT 
collaboration (Mattox et al., in preparation). 
Radio observations were done at the Mets\"ahovi Radio Observatory 
 (Ter\"asranta et al. 1998).

\placefigure{fig1}

\subsection {Multi-wavelength light curves}

The campaign began during a pronounced high amplitude flare recorded 
simultaneously by \sax and \whipple (Maraschi et al. 1999), 
and the high activity of Mrk~421 continued throughout the campaign (Fig.~1). 
The 2 -- 10~keV flux at the beginning of the \asca observation 
was $1.2 \times 10^{-10}$~erg~cm$^{-2}$~s$^{-1}$ 
and increased up to $5.0 \times 10^{-10}$~erg~cm$^{-2}$~s$^{-1}$ 
at the maximum.
 Thanks to the continuous coverage with \asca, the X--ray
variations  are fully resolved {\it with no gaps}. Many ``flares'' are
seen  clearly in the \asca data, superimposed on a general increasing
trend.  The \euve light 
curve also shows a series of flares, less well-defined because of 
lower amplitude and lower signal-to-noise ratio but well correlated 
in time with the X--ray flares.  The optical data for the period 
24 -- 30 April indicate an R-band flux with mean value 12.6~mag, 
and variability of less than 0.1~mag about this mean.  The high
frequency radio flux did not changed within statistics; the mean fluxes 
during the Mets\"ahovi observations are  0.5~Jy for 22~GHz  (20-26 April) 
and 0.6~Jy for 37 GHz (18-21 April).

In Mrk~421, the EUV through X--ray emission comprises the 
high energy part of the synchrotron component. Comparison of the 
data from \asca, \euve and \rxte indicates that the variability 
amplitude in the synchrotron component increases with photon energy. 
To show this quantitatively, 
we computed the fractional RMS variability including the general, 
increasing trend.
The
calculated RMS variability during \asca campaign increases with energy,
from $0.158
\pm 0.007$ in the \euve light curve, to $0.195 \pm 0.0004$ in \asca (0.5
-- 7~keV), to
$0.385
\pm 0.001$ and $0.67 \pm 0.03$ for 8 -- 16~keV and 16 -- 40~keV in the
\rxte PCA data, respectively. This energy dependence is clear
in the multi-epoch multi-frequency spectrum of Mrk~421 shown in
Fig.~2, where the largest changes in the synchrotron component occur
at the highest frequencies.

\placefigure{fig2}

The complete flare recorded both in X--rays by \sax and in TeV
$\gamma$--rays by the Whipple telescope demonstrates the correlation
between X--ray and TeV $\gamma$--ray emission on the time scale of hours
(Maraschi et al. 1999). The TeV flux also follows the general week-long
rise of the \asca flux ( Fig.~10 of Takahashi et
al. 1999). Assuming the cross-calibration among the three
TeV telescopes is accurate, the relative TeV intensities
at nearby times indicate that large intra-day variations are occuring
at these high energies.  Fractional TeV variability is respectively 
$0.30 \pm 0.07$ for the Whipple data and $0.45
\pm 0.08$ for the HEGRA data.  
Note that, for the Whipple data, 
$F_{var} = 0.43 \pm 0.05$ during the whole campaign.
These values are  similar to the values for the hard X--ray
data above 8~keV, consistent with the scenario that both TeV and hard
X-ray emissons are due to the same electron distribution.

In the subsequent discussion, we divided the \asca light curve into 
segments as marked in Fig.~3, with one flare per segment.  
In the \asca light curve, most of flares are characterized by nearly
equal rise and decay times, i.e., by a nearly symmetric time profile.
The time constants from the bottom to the top are also similar
from one flare to the next, $\tau \sim 0.5$~day.

\placefigure{fig3}

\subsection{Time series analysis}

The long, uninterrupted \asca light curve gives an excellent 
opportunity to study the temporal characteristics of the X--ray 
variability.
We therefore calculated the  first-order {\it normalized}  structure
function (SF) at time  lag $\tau$, which is defined as
$SF(\tau) = \frac{1}{N} \sum [(F(t) - F(t+\tau))/F_{av} ]^{2}$,
where $F(t)$ is the count rate and $F_{av}$ is the average count rate. 
$N$ is  the total number of data pairs used in the calculation. The
SF is closely related to the power density spectrum (PDS)
distribution (its slope, $\beta_{SF}$, is one less than the PDS slope) 
but is more suitable when the data sampling is uneven.
The SF calculated from the 2 -- 7.5~keV \asca light curve 
is shown in Fig.~4. It has a relatively steep slope at high temporal 
frequencies (short time scales), $\beta_{SF} \sim 1.2$ (corresponding to 
$\beta_{PDS}\sim 2.2$).  This is at the steep end of the slopes seen 
in other types of mass-accreting black hole systems, such as Seyfert 
galaxies (Hayashida et al. 1998).

A sharp break in the structure function of Mrk~421 occurs at $\sim 0.5$ 
days, to a flatter slope at low temporal frequencies (long time scales), 
$\beta_{SF} \sim 0.3$ ($\beta_{PDS} \sim 1.3$).  This is the first 
report of a clear turnover in the structure function of a blazar X--ray 
light curve, and it suggests there is at least one characteristic time 
scale for the variability of Mrk~421.  As shown in Fig.~4, the same 
flat slope at longer lag is seen in the structure function for the 
\rxte All-Sky Monitor (ASM) light curve (2 -- 10~keV), thus confirming 
the reality of the turnover. In order for the total variability power 
not to diverge, the structure function must flatten still further at 
longer time scales;  we infer this from the ASM data to be at $\gtrsim 
100$~days.
The shape of the structure function is roughly independent of energy from 
0.07 -- 7.5~keV, as can be seen from the comparison of structure functions
for the \asca data and the \euve data.

\placefigure{fig4}

\subsection{Spectral Variabilities and Time Lags}

The normalized light curves obtained with the SIS in two energy bands
(0.5 -- 1~keV and 3 -- 7.5~keV) are shown in Figs.~3a and 3b,
respectively. In general, the amplitude of the variability is larger in
hard X--rays than in soft X--rays. As can be seen from the hardness ratio
(Fig.~3c), the spectrum becomes harder when the source gets brighter.
Generally the spectral changes during individual flares are complex and rapid
but in some cases, the hardness ratio remains almost constant. In the
flare \#8, there is a plateau in the soft X--ray band, which is rarely
seen. According to detailed
analysis of the spectral variability (Kataoka 2000b), flare \#1 shows
clockwise hysteresis in the correlation between flux and photon index,
similar to that reported by Takahashi et al. (1996a).
 In contrast, we found anti-clockwise hysteresis in
flare \#7.
In order to quantify the lag, we used the discrete correlation function
(DCF; Edelson \& Krolik 1988). First, we calculated the DCF for the
\asca data comparing each of several energy bands to the 4 -- 7.5~keV
band using all data.
 Our results are consistent with zero-lag. We find the same
result even using the \euve data, which offers the largest separation
in energy from the hard \asca band.  However, inspection of X--ray
light curves at different energies does suggest some energy dependence
of the flare shape. We therefore calculated the lags separately for
each flare (as marked in Fig.~3) to see if individual flares have
different behavior.

Fig.~3d shows the time lag between 0.5 -- 1~keV photons (the softest
\asca band) and 4.0 -- 7.5~keV photons, calculated for each
flare. Errors of lags are determined from the quadratic sum of the
uncertainty in determining the maximum of the DCF distribution and the
results of MonteCarlo calculation which simulate the effect of time
windows (Kataoka 2000b). There are both soft (positive) and hard
(negative) lags, with some flares consistent with zero.  The soft lag
for flare \#1 is $2000\pm 710$~s for 0.5 -- 1.0~keV photons, which is
roughly half that found in previous monitoring in 1994 (Takahashi et
al. 1996). In contrast, the lag calculated for flare \#7 is negative
(a hard lag), $-3400 \pm 980$~s. These lags may not be real, as it is not
possible to reliably determine lags which are much smaller than an
orbit. Systematic effects could also produce these lags.

\section {Discussion}

The present multi-wavelength monitoring campaign showed that Mrk~421 
is much more variable, and that the variability, and 
its energy dependence, 
is more complex than anticipated.  The object 
flares daily (or perhaps even more often), and the correlation
of variations in the X--ray and TeV energy bands is confirmed, 
supporting the idea that the same electron distribution in the same
physical region is responsible for the emission in both energy bands.  
The 2 -- 7~keV energy index obtained with 
\asca ranges from $\alpha = 1.4$ to 1.8. The energy ranges covered 
by \asca and \rxte are thus at or above the peak in the $\nu$F$\nu$ 
spectrum. The increasing amplitude of variability at the higher 
X--ray energies is therefore consistent with the expectations for 
the synchrotron process, specifically that the higher energy electrons 
lose their energy more rapidly.

The X--ray variability of Mrk~421 is very similar to the intra-day
variability observed in the optical light curves of flat spectrum
radio quasars (FSRQs), for which the high-energy component peaks in the GeV 
$\gamma$--ray energy band (Wagner et al.  1996).  For synchrotron 
radiation due to a single population of relativistic electrons with 
a broken power law distribution of index $\gamma_{el}$ and a break at 
$\gamma_{peak}$, the correspondign  frequency for $\gamma_{peak}$ is
(as measured in the observer's frame) is $\nu_{sync} = 3.7 \times
10^{6} \gamma_{peak}^{2} B\frac{\delta}{(1+z)}$~Hz. Here
$B$ is magnetic field in Gauss, measured in the comoving frame, and
$\delta $ is the Doppler ``beaming'' factor defined as $\delta =
\Gamma_{j}^{-1} (1 - \beta \cos \theta)^{-1}$, where $\Gamma_{j}$ is the 
Lorentz factor of the jet, $\beta = v/c$, and $\theta$ the angle 
of the jet to the line of sight.  

Several analyses have shown that $\gamma_{peak}$ is lower for FSRQs 
($10^3$ -- $10^4$) and higher for HBLs ($\sim 10^5$ -- $10^6$), as 
is the case for Mrk~421 (Sambruna et al. 1996;  Kubo et al. 1998;  
Fossati et al. 1998).  Since the synchrotron emission due to the 
electrons with $\gamma_{peak}$ corresponds to the eV (or sub-eV) 
range for FSRQs and the keV range for Mrk~421, the rapid, large-amplitude 
variability observed above $\gamma_{peak}$ is likely the result of 
the rapid change of the electron distribution, in the regime where 
acceleration and cooling are approximately balanced 
(e.g., Inoue \& Takahara 1996;  Kirk et al. 1998).
Indeed, the structure function of Mrk~421 derived from the
X--ray light curve is very similar to the structure functions 
found in optical and radio studies of FSRQs, with a break at a 
characteristic time scale of one day (Wagner \& Witzel 1995;  
Wagner et al. 1996).

The \asca light curve constrains the dominant variability time scales
to be less than one day. The quasi-symmetric flare profiles
imply that $\tau_{acc}$ and $\tau_{cool}$ are shorter than the source
light-crossing time $\tau_{crs}$, because faster time scales will
always be smoothed out by $\tau_{crs}$ (Chiaberge \& Ghisellini 1999;
Kataoka et al. 2000c).  A plateau also occurs when the time scale for 
injecting energetic electrons, $\tau_{inj}$, is longer than 
$\tau_{cool}$. In this situation, emission from individual slices does 
not fade as quickly after an increase, hence the observation of a 
plateau in the light curve occurs when the entire volume is radiating. 
We do observe a plateau in the last flare (segment \#8), which is 
much more obvious at low energies, where the cooling time is longer.  

A simple conceptual scenario for multiple flares is that a cloud, or a 
``blob'' of 
plasma passes through the region where shock fronts are formed and
electrons are accelerated. If we assume the opening angle of the jet
is proportional to $1/\Gamma_{j}$, where the size of the blob at 
distance $D$ from the base of the jet can be expressed as 
$R \sim D/\Gamma_{j}$. With this, the
characteristic time scale of $\tau \sim 0.5$~days (from the structure
function analysis) implies that the emission region has size $R = c
\tau \Gamma_{j}$ $\sim 10^{16}$~cm at distance $D \sim 10^{17}$~cm from
the black hole, if we assume that $\Gamma_{j} \sim 10$.  
Further discussion
of the characteristic time scale   and a comparison with other TeV
blazars and Seyfert galaxies is presented elsewhere (Kataoka et al. 2000a).

If the detection of both positive and negative lags are real, it
implies that cooling is sometimes but not always the dominant process,
and that acceleration is also important. The (familiar) soft lag can
be explained by the fact that $\tau_{cool}$ is shorter for higher
energy electrons ($\tau_{cool} \propto
(1+u_{soft}/u_{B})^{-1}B^{-2}\gamma^{-1}\delta^{-1}$, where $u_{B}$
and $u_{soft}$ are the energy densities of the magnetic field and the
soft photons to be upscattered, respectively.  When high energy
electrons are injected into the emitting region and then cool
radiatively, the soft photons lag the hard, by a time roughly equal to
the cooling time at the soft energy (Takahashi et al. 1996).
Conversely, the suggested hard lag can be explained, at least
qualitatively, if the acceleration time is sufficiently long, $\tau_{acc} \sim
\tau_{cool}$, electrons appear first at low energies and gradually
build up at higher energies (Georganopoulos \& Marscher 1998).

The relatively small lags,
also mean $\tau_{acc} \sim \tau_{cool}$ at X--ray energies, at least 
during the present observations.  This is consistent with the fact
that the X--ray band is emitted by the highest energy regime of the
electron distribution, where $\tau_{acc}$ is
approximately balanced with $\tau_{cool}$.  According to the shock 
acceleration scenario, $\tau_{acc}$ depends on energy in an opposite sense to
$\tau_{cool}$. It follows $\tau_{acc} (\gamma) \propto \gamma$ for 
diffusive acceleration with a `gyro-Bohm' process
(e.g. Inoue \& Takahara 1996) or $\tau_{acc} (\gamma) \propto \gamma^{1/3}$ 
for fully developed Kolmogorov turbulence. Similar discussions have been
made by several authors (e.g. Sambruna 1999).

At a minimum, the present observations underline the need for detailed
time-dependent synchrotron models, in which the time scales
for electron acceleration and injection, and for radiative cooling and
escape, are free to vary (e.g., Kirk et al. 1998).  Depending on the 
relative values of these time scales, a variety of behaviors can be 
expected in the light curves.
The fact that plateaus are rarely seen implies the dominant time scale
is similar to the light-crossing time, which in term supports the idea
the the size of the emitting region is determined by diffusion and cooling.

\acknowledgments
CMU acknowledges support from NASA grant NAG5-3313.  GM acknowledges NASA 
grants to USRA and University of Maryland.

\clearpage

\figcaption[Fig1-ver4.eps]{Multiwavelength light curve for the 1998
campaign for Mkn~421. The \asca light curve is extracted by
integrating photons within $1.0'$ and $2.6'$ for SIS0 and SIS1,
respectively. The small radius for the SIS0 is to avoid the saturation
due to the limit of telemetry. The \euve count rate is extracted from
one orbit in a $12'$ aperture. The Whipple data are limited to be that
taken at elevations $>$ 55 degrees, so that the energy threshold and
sensitivity is similar for all runs. Each light curve is normalized to
its mean intensity. For TeV data, light curves from three telescopes
are first converted to Crab unit and then the combined light curve is
normalized.
\label{fig1}}

\figcaption[Fig2-ver4.eps]{Multi-frequency SED of Mrk~421 during
the 1998 campaign. Filled symbols are simultaneous data obtained
during the campaign. Different symbols correspond to different 
regions defined in Fig.1.  Flux in 22 GHz, 37 GHz and R-band are averaged
values during the campaign. The box from 0.07 -- 0.2 keV 
is the range of
the flux observed by the \euve during the campaign.  Butterfly-shape
boxes in TeV regions are from simultaneous HEGRA observations.
The EGRET data are from Macomb et
al. (1996) and other data are from the NED data base. Lines are predictions
by an one-zone SSC model for the steady state emission (Kataoka 2000). 
We fixed $B$ to be 
0.13 Gauss and $\delta$ to be 14.  $\gamma_{max}$ and electron 
normalization (cm$^{-3}$ s$^{-1}$) are (1.6$\times$10$^{5}$,
7.4$\times$10$^{-6}$) for the quiescent state and (2.5$\times$10$^{5}$,
5$\times$10$^{-5}$) for the high state.
\label{fig2}}

\figcaption[Fig-Lag-ver2.eps]{ Detailed time history of Mrk~421
emission obtained from \asca. (a) Normalized count rate in 0.5 -- 1.0~keV, (b)
3 -- 7.5~keV band, (c) hardness ratio of count rates, 
defined as (3 -- 7.5~keV)/(0.5 -- 1.0~keV), 
and (d) Time lag of photons of 0.5 -- 1~keV band from 3 -- 7.5~keV band
calculated from the DCF. Horizontal bar of each point
shows the coverage of data used in the calculation. 
\label{fig3}} 

\figcaption[SF_ver2.eps]{Structure function derived from the light
curve in the 2 -- 7.5~keV and 0.5 -- 2~keV bands with \asca (circles) and in
0.07 -- 0.2~keV band with \euve (filled triangles). Open triangles show the
results from ASM data in 2 -- 10~keV band accumulated from 1996 January 6 to 
1999 July 29, but normalized to the mean count rate during the present campaign.
\label{fig4}}

\clearpage 

\plotone{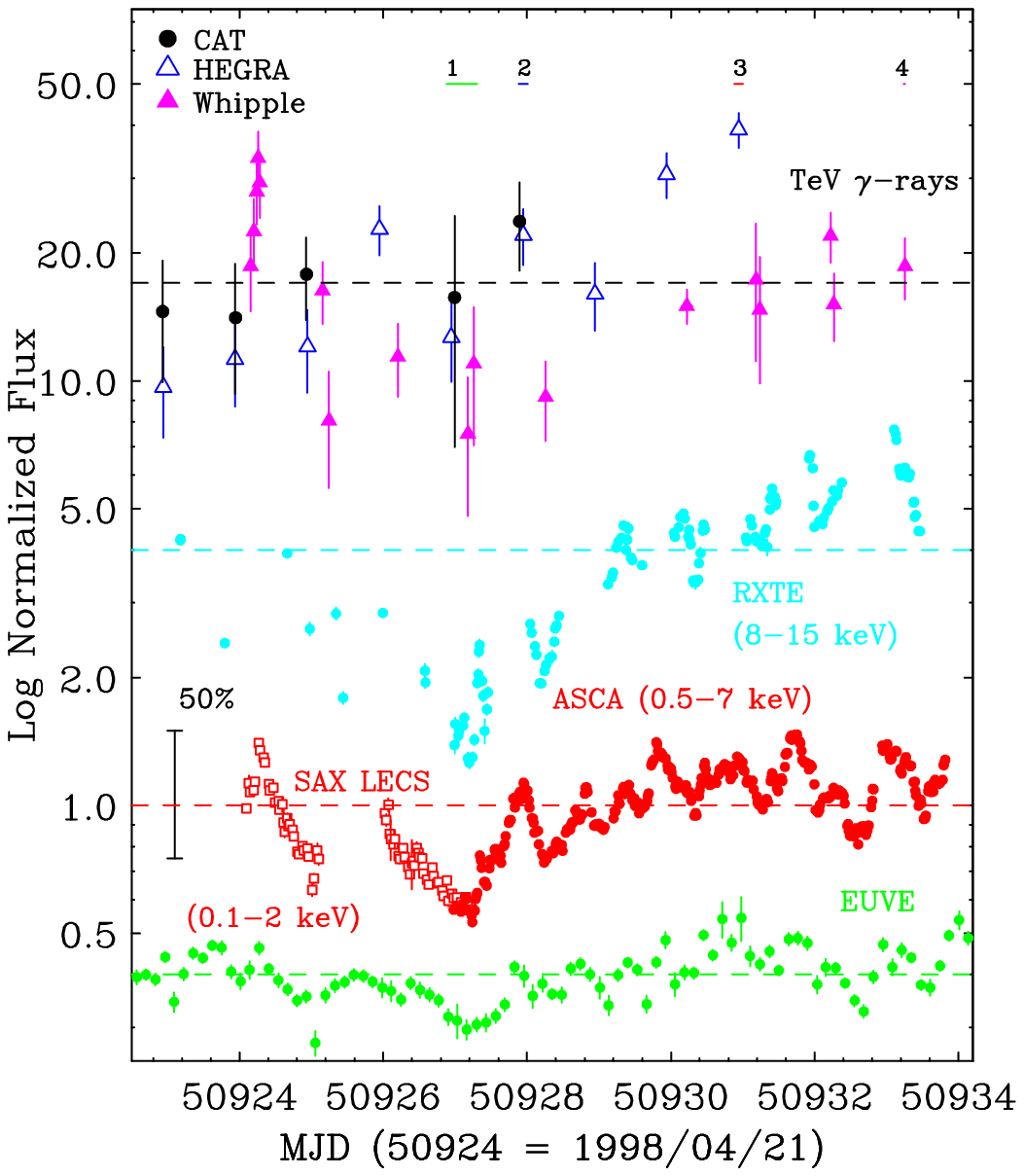}

\clearpage 

\plotone{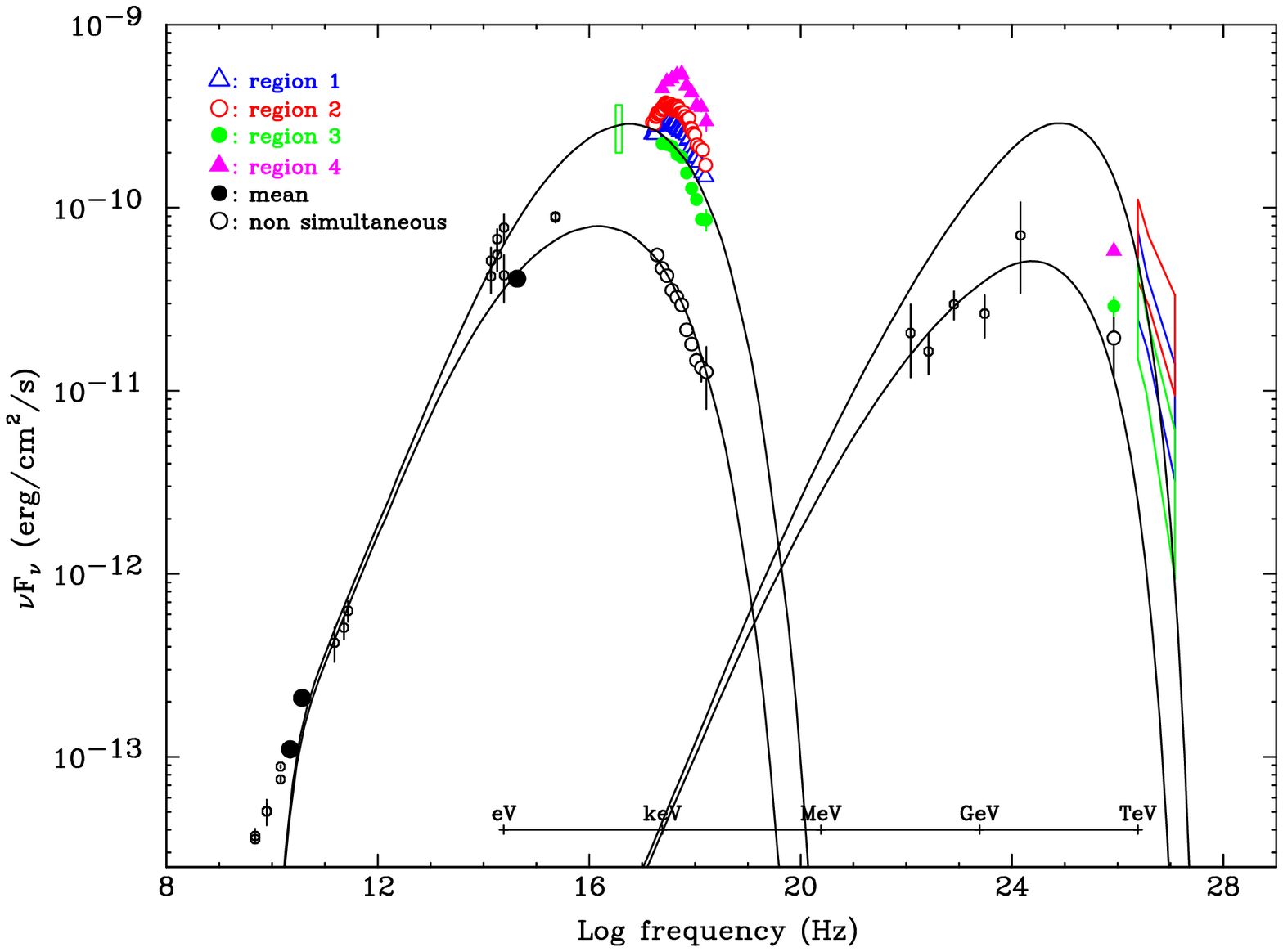}

\clearpage 

\plotone{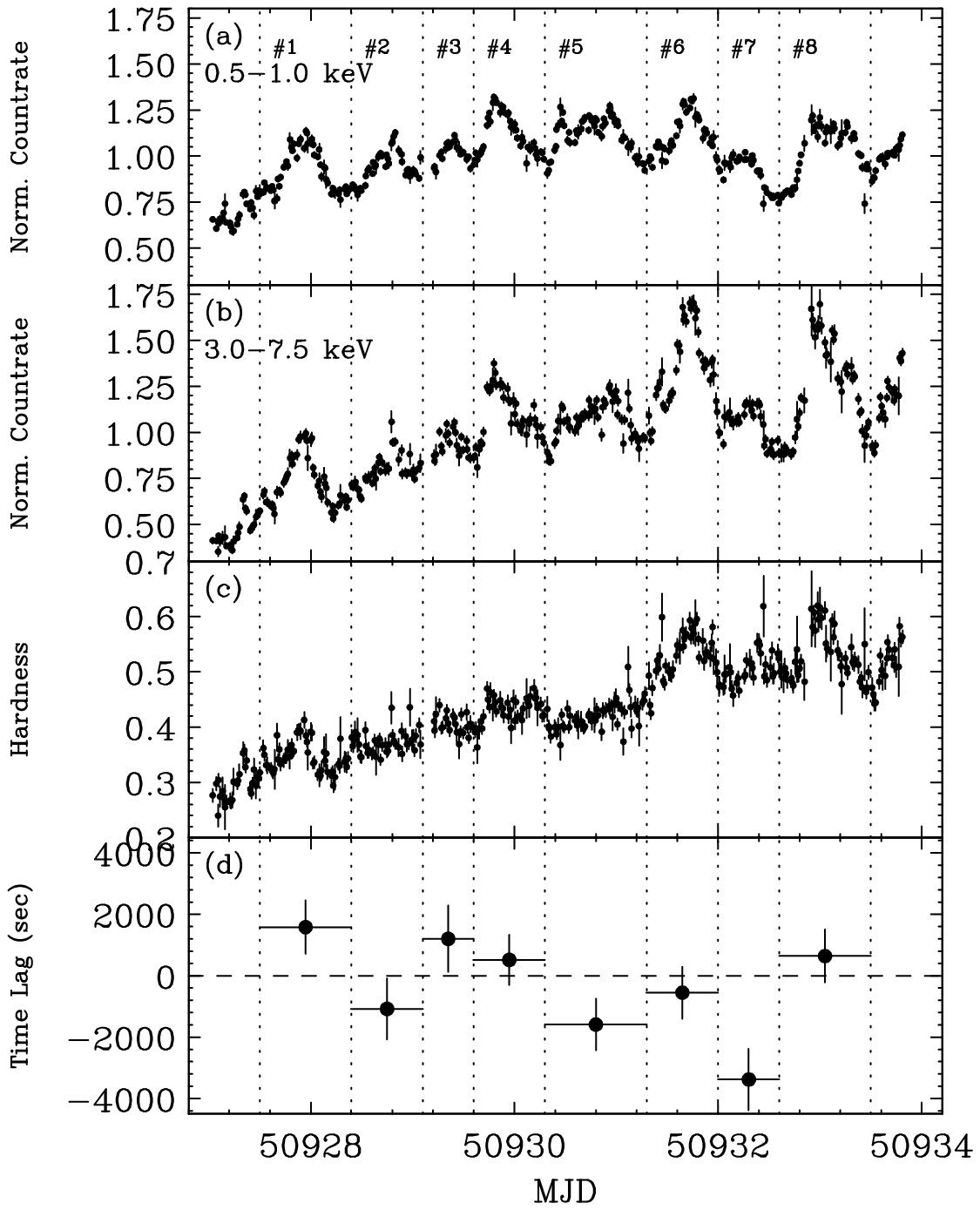}

\clearpage 

\plotone{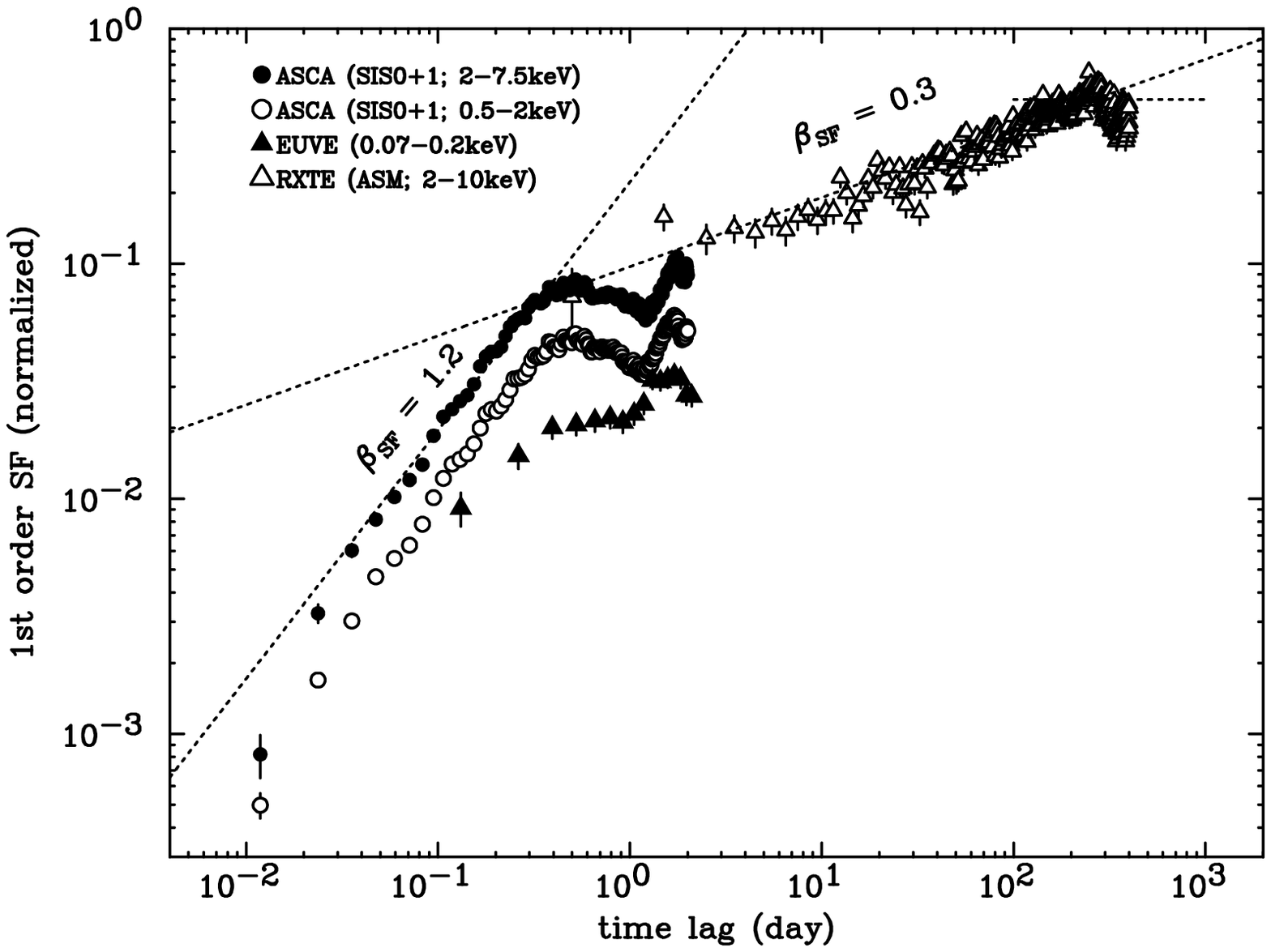}

\end{document}